\documentclass[electronic]{vgtc}             

\usepackage{booktabs} 
\graphicspath{{figures/}{pictures/}{images/}{./}} 

\usepackage{times}                     

\usepackage{tabu}                      
\usepackage{booktabs}                  
\usepackage{lipsum}                    
\usepackage{mwe}                       

\usepackage{tabularx}
\usepackage{array}

\usepackage{balance}

\usepackage{mathptmx}                  
\usepackage{amsmath}                   

\onlineid{2641}

\vgtccategory{Research}

\vgtcinsertpkg

\title{GaussAnything: Semantic Intent-Driven Refinement\\ of Evolving Gaussian Scenes for Standalone VR}

\author{Dmitrii Maliukov$^{1}$\thanks{e-mail: Dmitrii.Maliukov@skoltech.ru. Equal contribution.} %
\and Timofei Kozlov$^{1}$\thanks{e-mail: Timofei.Kozlov@skoltech.ru. Equal contribution.} %
\and Dmitrii Plotnikov$^{2}$\thanks{e-mail: Dmitrii.Plotnikov2@skoltech.ru} %
\and Miguel Altamirano Cabrera$^{1}$\thanks{e-mail: m.altamirano@skoltech.ru} %
\and Dzmitry Tsetserukou$^{1}$\thanks{e-mail: d.tsetserukou@skoltech.ru} 
}
\affiliation{\scriptsize 
$^{1}$ Intelligent Space Robotics Laboratory, Skolkovo Institute of Science and Technology, Moscow, Russian Federation \\
$^{2}$ NLP Research Center, Moscow, Russian Federation}

\teaser{
  \centering
  \makebox[\textwidth][c]{%
    \includegraphics[width=1.09\textwidth]{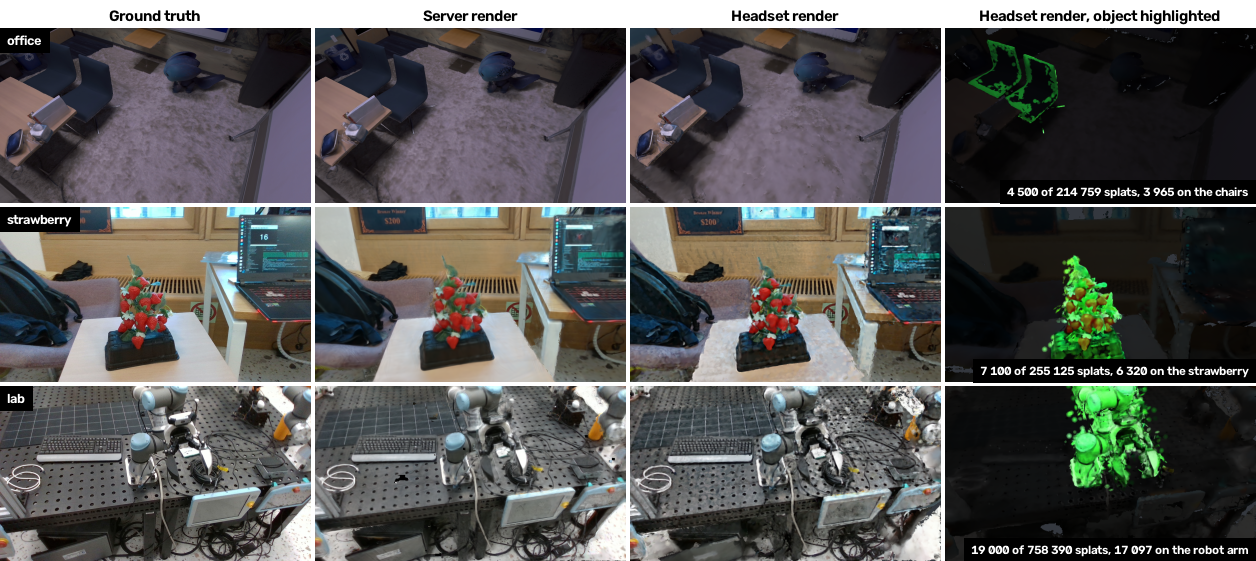}%
  }
  \vspace{-5mm}
  \caption{\textbf{GaussAnything} streams an evolving semantic Gaussian model and a TSDF-derived mesh
    to a standalone headset that renders stereo views
    on device under a fixed Gaussian budget. Columns: ground truth, server
    render, headset render, and headset render after a semantic query
    concentrates the budget on the selected object with optional highlighting.}
    \vspace{-2mm}
  \label{fig:teaser}
}

\abstract{
    Deploying reconstructed 3D environments on standalone VR headsets is constrained by limited compute and memory, and conventional level-of-detail policies optimize for visibility without accounting for the user's explicit inspection intent. We present \textbf{GaussAnything}, a native OpenXR system for intent-conditioned reallocation and progressive publication of evolving semantic Gaussian+SDF scenes. GaussAnything resolves class- or instance-level queries to persistent 3D objects and reallocates a fixed Gaussian resident budget toward the selected object while retaining global context, applying incremental, stable-identity updates coordinated with the TSDF-derived mesh through a source-epoch mechanism. Across eight scenes, an object query concentrates 88--90\% of the fixed client budget onto the queried object without enlarging it, on-device rendering reproduces the host render to within a small margin (up to 36.7~dB), and the standalone client renders each stereo frame at a steady-state GPU cost of roughly 10~ms---within the frame budget of standard standalone panels.
} 


\begin{document}


\firstsection{Introduction}

\maketitle

Recent advances in 3D Gaussian Splatting (3DGS) have made photorealistic scene representations increasingly practical for interactive rendering~\cite{kerbl2023gaussians}. Combined with online RGB-D reconstruction and SLAM, Gaussian representations can now be constructed and updated while an environment is being observed~\cite{splatam, gsslam}. Integrating a signed distance field (SDF)-based reconstruction framework creates new opportunities for immersive inspection: instead of receiving a remotely rendered video stream, a VR client can maintain a local representation of the reconstructed environment and render low-latency stereo views from its latest tracked pose. The user can therefore explore the scene independently of the initial camera trajectory while the underlying representation continues to evolve.

This setup introduces a unique challenge that does not occur when rendering static Gaussian scenes. As the reconstruction continuously evolves, updates constantly add, modify, and remove scene content. Applying these updates in virtual reality immediately can trigger sudden bursts of GPU workloads and abrupt visual artifacts. Conversely, delaying them preserves display continuity but results in an increasingly stale representation.

Importantly, maintaining a steady display frame rate does not automatically guarantee temporal stability; a client might meet its rendering deadlines but still expose massive model changes in a single frame. On the other hand, aggressively suppressing updates yields a stable but outdated world. Therefore, an immersive client must carefully balance the high-frequency display loop with the lower-frequency evolution of the scene representation.

Semantic interaction introduces a third requirement. During immersive inspection, not all scene content is equally relevant to the user's current task. Existing Gaussian methods have demonstrated view-dependent, perceptually guided, and semantic level-of-detail strategies~\cite{progs,srbf,sage,sggs}, while semantic Gaussian representations enable object localization, querying, and scene understanding~\cite{langsplat,online_language_splatting,objectgs}. Prior work demonstrates that Gaussian representations can efficiently allocate density to complex areas while embedding persistent semantic information directly into the scene. However, explicit semantic intent can also \emph{change} the desired representation during exploration. For instance, when a user selects a specific object for closer inspection, reallocating the limited client-side memory budget toward this target forces the immediate replacement of thousands of resident primitives. Enforcing immediate refinement minimizes the delay between user intent and high-quality rendering, yet triggers noticeable visual artifacts; staging it too conservatively reduces disruption but delays the requested detail. Thus, semantic refinement is not only a spatial resource allocation problem, but also a temporal publication problem.

We address this problem with \textbf{GaussAnything}, a framework for intent-conditioned refinement and publication of evolving semantic Gaussian scenes. GaussAnything maintains persistent semantic object IDs and interprets a class- or instance-level selection as an explicit representation objective. Under a fixed client Gaussian budget, it reserves additional appearance detail for the selected object while retaining a background context allocation. Crucially, the resulting change in the desired representation is not required to become visible immediately. Instead, GaussAnything decouples network reception from display-time publication, progressively integrating updates based on client workload and view-conditioned visibility. Latest-value coalescing discards superseded updates, freshness-aware servicing prevents indefinite deferral, and source-epoch coordination maintains consistency between the evolving Gaussian appearance and the supporting surface geometry.

This formulation separates three interacting temporal loops. The \emph{display loop} determines whether locally rendered stereo frames satisfy the immersive display deadline. The \emph{model loop} determines how quickly newly reconstructed scene state becomes visible without producing disruptive model-update transitions. The \emph{intent loop} spans the interval between an explicit semantic request and the availability of sufficient representation quality for the selected object. Optimizing any one loop in isolation can negatively affect another: maximizing publication throughput can increase visible disruption, minimizing change can increase staleness, and immediately prioritizing a selected object can produce a large representation transition. GaussAnything treats these objectives jointly while keeping head-coupled rendering local to the immersive client.

We implement GaussAnything in an end-to-end RGB-D reconstruction system: a host maintains an evolving hybrid GS+TSDF representation and persistent object-level semantic state, while a native OpenXR client maintains and renders the received scene model. This architecture preserves independent, head-tracked stereo rendering. Reconstruction, semantic inference, and network delivery proceed asynchronously, ensuring they do not block or dictate the viewpoint used for each displayed frame. 

 We evaluate GaussAnything on eight indoor scenes---four synthetic Replica rooms and four RGB-D captures---each reconstructed by an external system used as a black box. We measure the full delivery chain by comparing the ground-truth frame, the host (server) render, and the on-device headset render with standard image metrics (PSNR, SSIM, LPIPS); the fraction of a fixed client Gaussian budget that a semantic object query redirects onto the selected object; the sensitivity of headset fidelity to that budget; and standalone rendering cost reported as per-frame GPU/CPU time. Reconstruction quality is reported only as the reference ceiling, and resident Gaussian count is treated as a representation-state metric rather than visual-quality evidence.
 
\emph{Our work makes three main contributions.} First, we formulate semantic inspection of evolving Gaussian scenes as a joint spatiotemporal problem, exposing the interaction between display continuity, model freshness, and semantic intent-to-quality. Second, we introduce GaussAnything, which uses explicit semantic object selection to reallocate a fixed Gaussian budget and progressively publishes updates to prevent visual disruption. Finally, we evaluate the system across eight scenes, quantifying the transmission fidelity from host to headset, the dynamic reallocation of a fixed budget under semantic queries, the budget required by a resource-constrained client to maintain viewpoint-local fidelity, and the on-device rendering cost.

\begin{figure*}[t]
      \centering
      \includegraphics[width=0.85\textwidth]{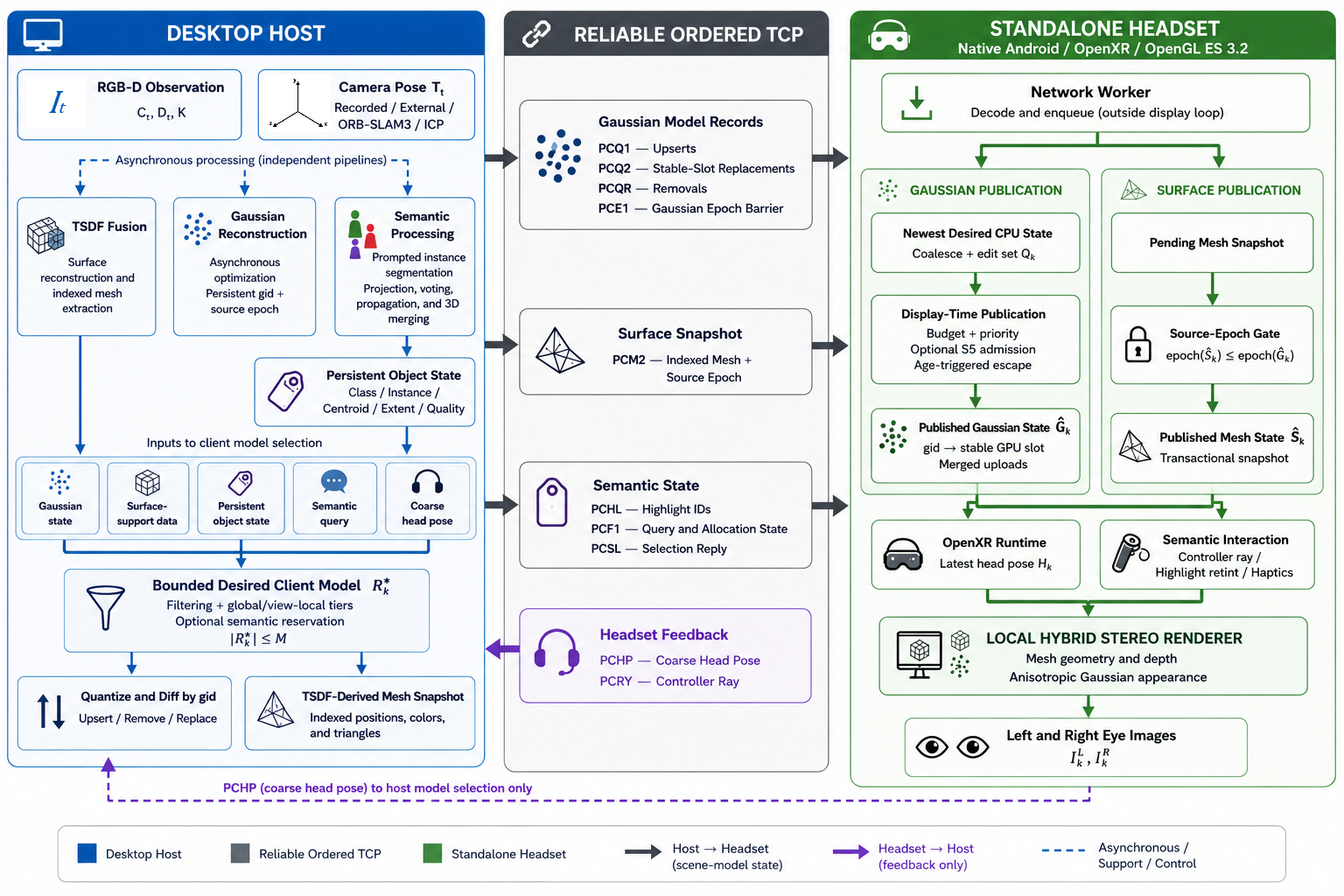}
      \vspace{-5mm}
      \caption{GaussAnything architecture. The host maintains an evolving semantic
      Gaussian+SDF model and transmits stable-ID updates to a standalone
      client, which separates desired and published state and locally
      renders stereo views in OpenXR.}
      \vspace{-4mm}
      \label{fig:architecture}
  \end{figure*}

\section{Related Work}

3D Gaussian Splatting (3DGS) combines high-quality novel-view synthesis with real-time rasterization~\cite{kerbl2023gaussians}. Subsequent systems such as SplaTAM~\cite{splatam}, Gaussian Splatting SLAM~\cite{gaussian_splatting_slam}, and GS-SLAM~\cite{gsslam} extended Gaussian representations to online mapping, incrementally creating, optimizing, and removing primitives as new observations arrive. GPS-SLAM \cite{gpsslam} and AnythingReality \cite{AR} utilize SDFs as a scaffold to guide Gaussian initialization, significantly increasing both reconstruction efficiency and frame rate. More recent methods additionally maintain semantic information during reconstruction, including Online Language Splatting~\cite{online_language_splatting} and S2GS~\cite{s2gs}. GaussAnything builds on these capabilities rather than proposing a new reconstruction or SLAM algorithm; our focus is downstream, on how a continuously changing reconstructed model is maintained and published by an independently rendering standalone VR client.

The size and rendering cost of Gaussian representations have motivated progressive transmission and adaptive level of detail. PRoGS~\cite{progs} orders Gaussians according to their contribution, LapisGS~\cite{lapisgs} organizes scenes into representation levels, and SRBF-Gaussian~\cite{srbf} combines viewport-aware pruning and multi-level representations for immersive rendering. Recent works stream continuously optimized GS models through snapshots and incremental updates, allowing clients to reconstruct and render changing scene state locally~\cite{streaming_realtime_gaussians}. These approaches address which content should be transmitted or retained under bandwidth, viewpoint, and compute constraints. GaussAnything instead emphasizes the client-side publication of mutations to an already resident, continuously evolving model. Persistent identities allow additions, removals, and replacements to update stable GPU state incrementally, separating receipt of a model revision from its integration into the rendered state.

Semantic Gaussian representations further support object-level localization, querying, and scene understanding. LangSplat~\cite{langsplat} embeds language-aligned features into 3D Gaussians, Online Language Splatting~\cite{online_language_splatting} performs online language mapping, and ObjectGS~\cite{objectgs} supports object-aware reconstruction and interaction. Semantic information has also been used to control Gaussian detail: SAGE~\cite{sage} and SGGS~\cite{sggs} adapt representation or level of detail according to semantic importance. Semantic selection and semantic level of detail are therefore not contributions of GaussAnything by themselves. Instead, GaussAnything leverages persistent state at both the class and instance levels as a control signal for a standalone client with fixed capacity. This allows object queries to deterministically dictate which appearance Gaussians allocate memory within the resident budget, while seamlessly preserving background context.

Immersive rendering imposes additional stereo frame-time and memory constraints. Recent systems adapt Gaussian rendering according to perceptual importance, viewpoint, gaze, or available resources~\cite{perceptual_3dgs_vr,srbf}, while VR-specific methods address Gaussian rendering artifacts and stereo or temporal efficiency~\cite{vrsplat,gsreuse}. More broadly, perceptually optimized progressive VR systems have shown that temporal consistency and the order in which representation detail becomes visible can influence immersive presentation. GaussAnything considers a related but distinct setting in which the representation itself remains mutable during exploration. Rather than treating adaptation only as rendering or level-of-detail selection, GaussAnything maintains stable client-side scene state and incrementally integrates model revisions while local OpenXR rendering continues.

While prior work addresses online Gaussian reconstruction, semantic representations, and progressive delivery individually, GaussAnything explores their intersection on standalone clients. Specifically, we investigate how to publish consistent, object-level updates from an evolving semantic Gaussian-SDF reconstruction within a bounded memory footprint. Consequently, our evaluation explicitly distinguishes the target model state from the currently resident client state, allowing us to separately track model convergence, update-induced visual artifacts, and real-time display performance.

\section{System Architecture}
\label{sec:system_architecture}

GaussAnything is an end-to-end distributed system for scene management and immersive rendering, where a desktop host continuously updates a hybrid 3D representation while a standalone VR client independently maintains and renders a bounded local model. The host leverages an upstream pipeline for RGB-D tracking, surface reconstruction, Gaussian optimization, and semantic association. Our system converts this evolving scene into a bounded Gaussian set and a TSDF-derived indexed mesh with a fixed Gaussian budget. Rather than receiving remotely rendered stereo images, the immersive client receives scene-model updates and generates stereo views locally using the latest OpenXR-tracked head pose. Reconstruction, semantic processing, network delivery, and display tracking therefore proceed asynchronously and can evolve at different rates.

\emph{Evolving scene representation.}
At reconstruction time $t$, the host receives an RGB-D observation
\begin{equation}
I_t = \left(C_t,D_t,K,T_t\right),
\label{eq:rgbd_observation}
\end{equation}
where $C_t$ and $D_t$ denote color and depth images, $K$ contains camera intrinsics, and $T_t\in SE(3)$ is the estimated camera pose. Depending on the configuration, poses are obtained from recorded trajectories or an external source such as ORB-SLAM3 or odometry. Controlled publication experiments use fixed or ground-truth trajectories to separate client-side publication behavior from tracking drift.

The host maintains an evolving Gaussian representation
\begin{equation}
\mathcal{G}_e = \{g_i\}_{i=1}^{N_e},
\label{eq:gaussian_scene}
\end{equation}
associated with source epoch $e$, together with a supporting surface representation $\mathcal{S}_e$. Each Gaussian stores position, anisotropic scale and orientation, opacity, color, observation support, a persistent identifier, and, when available, a semantic object identifier.

RGB-D observations are incrementally fused into a TSDF-derived surface, while Gaussian appearance is optimized asynchronously from recent observations. New reconstruction triggers may supersede older pending optimization work instead of requiring every input frame to produce a complete Gaussian update. The two representations serve complementary roles: Gaussians provide appearance and the mutable state used by residency and publication policies, whereas the surface provides stable geometry and depth for hybrid rendering and interaction. Crucially, GaussAnything does not introduce a new SLAM or reconstruction algorithm; these components produce the evolving scene state consumed by the immersive client.

\emph{Persistent semantic state.}
The host leverages an asynchronous semantic processing pipeline to enrich the 3D scene representation. Under this upstream pipeline, selected RGB frames are processed by a text-prompted instance-segmentation service; the resulting masks, class labels, and confidence values are tracked across sequential observations and projected into the evolving 3D space using camera geometry. Visible Gaussians accumulate these semantic observations over time, while temporal tracking and instance-merging mechanisms maintain persistent object identities across changing viewpoints.

\begin{figure*}[t]
    \centering
    \includegraphics[width=0.85\textwidth]{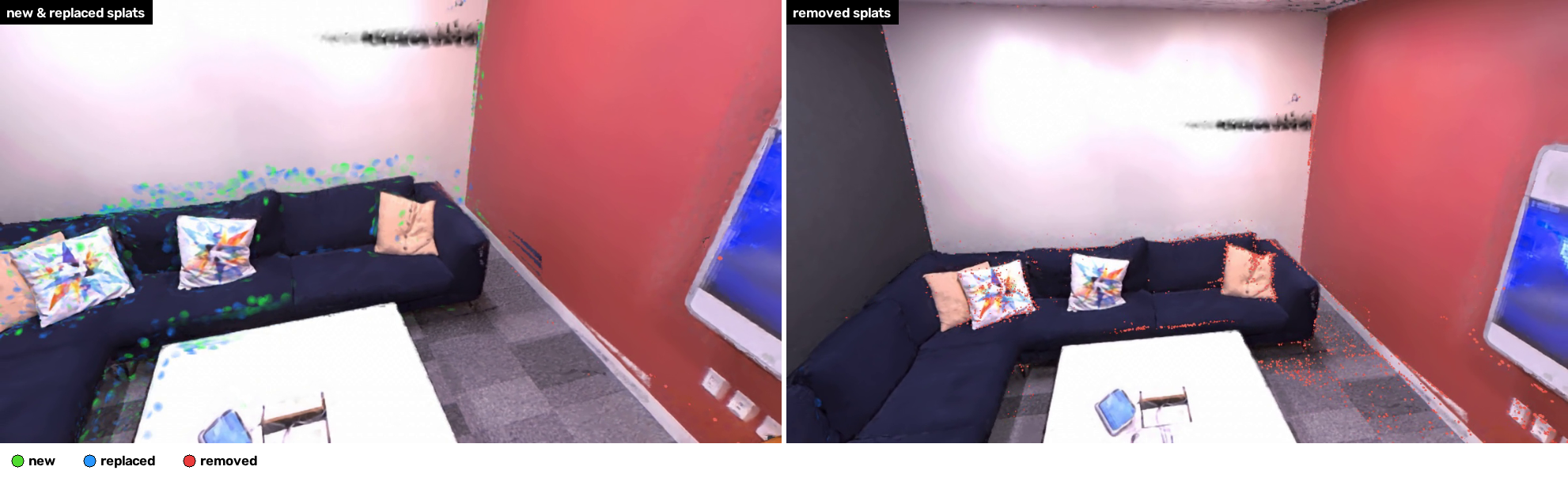}
    \vspace{-5mm}
    \caption{Live model-update visualization on the headset during a
    fly-through of \texttt{office2}. As the viewpoint moves, the client
    marks streamed Gaussians as newly added (green), view-aware replaced
    (blue), or removed (red), showing that the bounded resident model is
    continuously maintained under motion.}
    \vspace{-4mm}
    \label{fig:stream}
\end{figure*}

For each persistent object $o_j$, the host tracks its semantic class, associated Gaussian identifiers, centroid, spatial extent, and observation-quality metrics. A class-level user query can resolve to multiple persistent object instances, whereas an instance-level query isolates a single target object. These persistent identifiers are shared across both semantic highlighting and client representation selection mechanisms. Crucially, semantic carrier primitives used primarily for upstream object association are distinguished from appearance Gaussians and are strictly excluded from our measurements of client-side photometric residency.

\emph{Bounded client representation.}
The host can maintain substantially more eligible Gaussians than the
standalone client can keep resident. Before transmission, Gaussians are
filtered using reconstruction-dependent criteria such as opacity, scale,
observation support, and surface consistency. Let
$\mathcal{R}_k^{\star}$ denote the desired client resident set at time
$k$. Its size is bounded by
\begin{equation}
|\mathcal{R}_k^{\star}| \leq M,
\label{eq:resident_capacity}
\end{equation}
where $M$ is the configured client capacity.

Under ordinary view-dependent selection, the host combines persistent
global content with Gaussians relevant to the predicted headset view.
The client periodically returns a coarse head pose for host-side
selection, but this pose is never used for display-time tracking. Local
rendering always uses the latest pose supplied by the OpenXR runtime.

When semantic intent is active, the selected object set
$\mathcal{J}_k$ divides the eligible appearance Gaussians into a
selected-object set $\mathcal{G}_{\mathrm{obj}}$ and a background set
$\mathcal{G}_{\mathrm{bg}}$. Specifically,
$\mathcal{G}_{\mathrm{obj}}$ contains Gaussians whose persistent semantic
identity belongs to $\mathcal{J}_k$, while all remaining eligible
Gaussians form $\mathcal{G}_{\mathrm{bg}}$.

For reservation fraction $\rho$, the selected-object capacity is
\begin{equation}
M_{\mathrm{obj}}
=
\min\left(
|\mathcal{G}_{\mathrm{obj}}|,
\lfloor \rho M \rfloor
\right).
\label{eq:object_budget}
\end{equation}

The remaining capacity is filled from the background set using the
ordinary view- and quality-aware selection policy. Within each
partition, candidate Gaussians are ranked using reconstruction support
and view-dependent utility. Semantic intent therefore changes the
target resident composition without increasing the total client model
size. The implementation supports up to 400,000 stable Gaussian slots;
the controlled experiments use $M=200{,}000$.

\emph{Stable-identity scene-model transport.}
Selected Gaussians are quantized into compact records before
transmission. Persistent identifiers allow consecutive desired states
to be represented as insertions, removals, updates, and stable-slot
replacements rather than complete scene snapshots. Surface updates,
semantic state, and source-epoch information are transported alongside
the Gaussian appearance stream.

Transport uses a reliable ordered connection. Network reception
executes outside the OpenXR display loop and updates CPU-side desired
state and publication queues without directly modifying the visible GPU
representation. Receiving a newer logical model state and exposing that
state to the renderer are therefore distinct events.

At display frame $k$, the client locally renders
\begin{equation}
(\mathcal{I}_k^{L},\mathcal{I}_k^{R})
=
\mathcal{F}(
\widehat{\mathcal{G}}_k,
\widehat{\mathcal{S}}_k,
H_k).
\label{eq:local_rendering}
\end{equation}
Here, $\widehat{\mathcal{G}}_k$ and $\widehat{\mathcal{S}}_k$ are the
currently published Gaussian and surface states, $H_k$ is the latest
OpenXR-tracked head pose, and $\mathcal{I}_k^{L}$ and
$\mathcal{I}_k^{R}$ are the left- and right-eye images. The published
state may temporarily lag behind the newest desired state already
received from the host.

\emph{Surface transport.} The TSDF-derived surface is sent to the client as a compact indexed mesh. Unlike the continuously updated appearance stream, this coarse geometry proxy is largely static and is transmitted once (a few megabytes), with only incremental patches thereafter; its bandwidth and client-side cost are therefore negligible next to the Gaussian stream and impose essentially no additional load on the headset. Crucially, the client can rasterize this mesh immediately to obtain an approximate, gap-free geometry of the scene---providing depth and occlusion before, and independently of, the finer Gaussian appearance layer, onto which the Gaussians are then composited as they arrive.

\emph{Native immersive client.}
The client is implemented as a native Android OpenXR application using OpenGL ES~3.2. A preallocated Gaussian buffer provides stable GPU slots indexed by persistent Gaussian identifiers, allowing individual model mutations to be integrated without rebuilding the complete representation. Dirty stable-slot ranges are merged before upload to reduce the number of GPU buffer operations.

For each persistent identifier, the client maintains both the newest desired logical state and, when resident, its current published GPU state. Because reconstruction and semantic selection can continue while earlier changes are still queued, several received edits may refer to the same Gaussian. GaussAnything therefore applies latest-value coalescing: if a newer desired value arrives before an older pending value is published, the obsolete intermediate state is discarded. This allows the client to converge directly toward the most recent target representation.

The hybrid renderer first rasterizes the supporting surface to establish geometry and depth and then composites projected anisotropic Gaussians with the surface representation. Rendering occurs entirely on the headset using runtime-tracked stereo viewpoints. Controller geometry and interaction share the same local scene depth rather than relying on a remotely rendered image.

\begin{table}[t]
  \centering
  \caption{Reference quality of the source models, produced by
  AnythingReality and used here as a fixed input
  (not a contribution of this paper). PSNR/SSIM/LPIPS under the depth-mask
  convention on held-out eval renders; \#G = model size in Gaussians,
  the operative quantity for the fixed-budget client study. Top: 
  RealSense captures; bottom: synthetic Replica rooms.}
  \label{tab:recon}
  \begin{tabular}{l r r r r r}
    \toprule
    Scene & Frames & PSNR (dB) & SSIM & LPIPS & \#G \\
    \midrule
    \multicolumn{6}{l}{\emph{RealSense captures}}\\
    strawberry & 900  & 26.68 & 0.862 & 0.273 & 253k \\
    tables     & 1799 & 26.24 & 0.908 & 0.319 & 580k \\
    kitchen    & 1799 & 23.70 & 0.874 & 0.355 & 922k \\
    lab1       & 1001 & 22.66 & 0.750 & 0.307 & 758k \\
    \midrule
    \multicolumn{6}{l}{\emph{Synthetic (Replica)}}\\
    office0    & 2000 & 39.03 & 0.964 & 0.106 & 215k \\
    office1    & 2000 & 38.07 & 0.953 & 0.173 & 95k  \\
    office2    & 2000 & 32.28 & 0.934 & 0.168 & 417k \\
    room0      & 2000 & 31.71 & 0.917 & 0.122 & 892k \\
    \bottomrule
  \end{tabular}
  \vspace{-7mm}
\end{table}

\emph{Display-time model publication.}
GaussAnything maintains the newest desired client state separately from the state currently exposed to the renderer. Immediate publication applies stable-ID edits as soon as possible, whereas progressive publication modes integrate only a subset of pending edits during each display frame. Network arrival therefore does not imply immediate visibility.

Let $\mathcal{Q}_k$ denote the pending edit set at frame $k$ and $\mathcal{A}_k\subseteq\mathcal{Q}_k$ the edits selected for publication. The scheduler bounds client-side integration work while prioritizing edits according to reconstruction support, current view relevance, change magnitude, and age. More elaborate publication modes additionally use a heuristic head-conditioned change score when ordering and admitting updates. This score is used only as an internal scheduling signal and is not treated as a calibrated perceptual metric.

The work allowance adapts within predefined limits according to recent display slack. Accepted edits are integrated into their stable GPU slots before the subsequent stereo render, while unaccepted edits remain queued. A best-effort age-triggered escape allows limited service of sufficiently old edits even when they would otherwise remain delayed by the normal publication policy. This mechanism improves scheduler progress but does not impose a hard freshness or completion deadline.

GaussAnything consequently distinguishes several stages of model evolution: an edit can be available on the host, received by the client, present in the desired logical state, queued for publication, integrated into stable GPU state, and finally visible in a locally rendered frame. The evaluation records these stages separately rather than equating network delivery with displayed model freshness.

\emph{Hybrid source-epoch coordination.}
Gaussian appearance and supporting surface geometry can be generated asynchronously from the same evolving reconstruction. To reduce cross-representation ordering inconsistencies, both streams carry source epochs. The client tracks the highest Gaussian epoch whose required edits have been integrated and normally commits a newer surface state only when
\begin{equation}
\operatorname{epoch}
\bigl(\widehat{\mathcal{S}}_k\bigr)
\leq
\operatorname{epoch}
\bigl(\widehat{\mathcal{G}}_k\bigr).
\label{eq:epoch_constraint}
\end{equation}
This global barrier prevents the published surface state from normally advancing beyond the corresponding Gaussian state. It is a causal state-management mechanism rather than a claim of perceptual coherence.

\emph{Semantic interaction and observability.}
Class- and instance-level selections provide immediate acknowledgement through highlighting of the corresponding persistent object identifiers. Highlighting is independent from any associated change in resident Gaussian composition and can therefore become visible before the requested target resident state has been reached.

The system records query resolution, semantic target identity, initial and desired residency, first additional resident content, final target convergence, publication queue state, frame timing, and GPU integration behavior as separate events. This instrumentation makes it possible to distinguish interaction feedback, target-state progress, model publication, and local rendering behavior during evaluation.

Overall, GaussAnything separates four scene states that are commonly conflated in progressive or remote rendering pipelines: the evolving scene maintained by the host, the bounded representation desired for the client, the subset currently published into stable GPU state, and the stereo image rendered from that published state at the latest tracked pose. This separation allows reconstruction, semantic interaction, representation selection, network transport, model publication, and immersive rendering to proceed asynchronously while remaining explicitly observable and independently measurable.

\section{System Evaluation}
\label{sec:evaluation}

We evaluate GaussAnything as a publication-and-rendering system: given an evolving semantic Gaussian+SDF reconstruction from an existing system (AnythingReality, treated as a fixed black box), we study how accurately that model is delivered across the host-to-headset boundary and rendered on a standalone device. Scene reconstruction is not a contribution of this paper; its output quality serves only as the reference ceiling against which we measure the delivered on-device fidelity. The study is organized around four key questions. (i) How much of the source model's fidelity survives streaming and rendering on the headset? (i.e., how far does the delivered image depart from its upper-bound reconstruction?) (ii)  Does local rendering on the headset preserve the host's original appearance once the model is published over the network? (iii) Can explicit semantic intent redirect a fixed client-side Gaussian budget toward a queried object without enlarging the budget? (iv) Does the client sustain interactive stereo rendering while these operations run? We report the results of each stage separately rather than collapsing them into a single end-to-end score.

\subsection{Experimental Setup}

\emph{Scenes.} We evaluate our approach using eight indoor scenes across two distinct categories. The first consists of four synthetic Replica rooms  (office0, office1, office2, room0), each containing 2000 posed RGB-D frames. The second comprises four real-world RealSense captures (strawberry, tables, kitchen, lab1) of cluttered tabletops and robot workcells containing semantically queryable objects (e.g., a potted strawberry plant, a keyboard, and a UR robot arm). This pairing allows us to isolate reconstruction-limited behavior from capture- and sensor-limited artifacts.

\emph{Source models.} All scenes are reconstructed with AnythingReality
under a single fixed configuration, identical densification, keyframe, and TSDF settings, ORB-SLAM3 poses, and GES Gaussian training. Scenes are replayed in real time at 10 FPS without per-scene tuning. As noted, this reconstruction system is treated strictly as an unchanged black box; we report its output solely to establish the reference quality and the baseline model sizes for our client-side studies.

\begin{table*}[t]
  \centering
  \caption{Rendering fidelity across the GT $\rightarrow$ PC
  $\rightarrow$ HS chain, and standalone headset performance.
  GT-PC is the reference stage; PC-HS and GT-HS are the delivered on-device
  results. Conv.\ is
  convergence time after a far teleport; Flk.\ is post-convergence
  flicker. FPS is measured on the PICO~4 Ultra. }
   \label{tab:fidelity}
  \begin{tabular*}{\textwidth}{@{\extracolsep{\fill}} l r r r r r c c c @{}}
    \toprule
    & \multicolumn{1}{c}{GT-PC} & \multicolumn{3}{c}{PC-HS} & \multicolumn{1}{c}{GT-HS} & & & \\
    \cmidrule(lr){2-2}\cmidrule(lr){3-5}\cmidrule(lr){6-6}
    Scene & PSNR & PSNR & SSIM & LPIPS & PSNR & Conv.\,(s) & Flk. & FPS (mean/p05) \\
    \midrule
    \multicolumn{9}{l}{\emph{Synthetic (Replica)}}\\
    office0 & 37.59 & 36.67 & 0.705 & 0.277 & 32.37 & 0.92 & 0.000 & 77 / 39 \\
    office1 & 39.52 & 34.51 & 0.518 & 0.174 & 34.28 & 0.87 & 0.000 & 70 / 36 \\
    office2 & 32.64 & 21.90 & 0.771 & 0.252 & 26.93 & 0.99 & 0.073 & 66 / 45 \\
    room0   & 30.15 & 22.88 & 0.736 & 0.318 & 24.40 & 0.93 & 0.201 & 67 / 34 \\
    \midrule
    \multicolumn{9}{l}{\emph{RealSense captures}}\\
    strawberry & 26.25 & 24.70 & 0.871 & 0.215 & 19.61 & 0.94 & 0.002 & 85 / 49 \\
    tables     & 27.51 & 25.38 & 0.898 & 0.228 & 21.82 & 0.90 & 0.128 & 81 / 42 \\
    kitchen    & 21.64 & 27.71 & 0.825 & 0.304 & 20.88 & 1.05 & 0.017 & 78 / 39 \\
    lab1       & 22.10 & 19.37 & 0.728 & 0.310 & 16.70 & 1.63 & 0.095 & 80 / 40 \\
    \bottomrule
  \end{tabular*}
  \vspace{-4mm}
\end{table*}

\emph{Measurement chain.} Every fidelity measurement follows a standard pipeline.
First, we take the ground-truth (GT) dataset frame. Next, we generate the host free-view render from the same camera position on the server. Finally, we capture the corresponding headset render directly from the device. To ensure a clean capture, we use the headset's native viewport-lock path, which excludes controllers, HUD elements, and isolates the left eye. We eye-align the requested camera with the dataset camera, using a settle phase to re-capture the pose until the on-device poses stabilize. GT-referenced metrics zero out pixels without valid GT depth in both images, and the metric is averaged over the entire frame. However, host-vs-headset (PC-HS) comparisons require a different approach because the headset field of view (FOV) is wider than that of the dataset camera. For these pairs, we background-normalize both renders and calculate the metric exclusively over their content intersection.

\begin{table}[t]
  \centering
  \caption{Semantic reallocation of a fixed client budget. An object
  query concentrates the resident appearance Gaussians on the requested
  object without changing the total budget. $N_\mathrm{obj}$ is the number
  of resident Gaussians on the object; Share is $N_\mathrm{obj}$ divided
  by the budget.}
  \label{tab:semantic}
  \begin{tabular}{l l r r r}
    \toprule
    Scene & Object (query) & Model & Budget & $N_\mathrm{obj}$ (Share) \\
    \midrule
    strawberry & plant (class)     & 255k & 7{,}100  & 6{,}320 (89\%) \\
    office0    & chairs (inst.)    & 215k & 4{,}500  & 3{,}965 (88\%) \\
    lab1       & robot arm (inst.) & 758k & 19{,}000 & 17{,}097 (90\%) \\
    lab1       & keyboard (inst.)  & 758k & 23{,}700 & 21{,}245 (90\%) \\
    \bottomrule
  \end{tabular}
  \vspace{-4mm}
\end{table}

\emph{Protocol.} While evaluating our system we focused on four key dimensions. First, we assess temporal behavior by teleporting the view to a distant pose, returning to the target pose, and tracking how quickly the published model converges (quantified via SSIM against the settled frame); we then measure per-pixel flicker across the post-convergence captures. Second, to evaluate semantic reallocation, we issue an object query and compare the resident model’s behavior with and without semantic reservation at a fixed client capacity. Third, for the budget ablation, we replay each scene across client capacities ranging from 25k to 400k resident Gaussians, reporting the PC-HS PSNR. Finally, we evaluate reconstruction quality using the reconstruction system's evaluation script on the held-out evaluation renders of each run.

\emph{Hardware.} The host is a single RTX 5090; the client is a PICO 4 Ultra
running the native OpenXR / OpenGL ES viewer. All headset numbers are
measured on the physical device.

\subsection{Reference Model Quality}

Because reconstruction is external to this work, we report the quality of
the source models only as the reference ceiling for the delivery study
that follows. Tab.~\ref{tab:recon} lists the source-model PSNR/SSIM/LPIPS
under the depth-mask convention together with the model size in
Gaussians. These numbers are produced by AnythingReality
 and are not a contribution of this paper; they define
the upper bound that the streamed, on-device rendering is measured
against in Section 4.3. Quality tracks scene difficulty as expected
(synthetic rooms 31.7–39.0 dB, RealSense captures 22.7–26.7 dB), and
model size ranges from 95k to 922k Gaussians. The latter is the operative
figure for this paper: the full models routinely exceed a practical
standalone-headset residency budget, which motivates the fixed-budget
client study in Section 4.4.

\subsection{Cross-Boundary Rendering Fidelity}

Tab.~\ref{tab:fidelity} reports the three-stage chain GT → server →
headset. The GT-PC stage serves as the baseline, reflecting the inherent quality of the source model. Our primary contribution lies in the subsequent transmission and rendering stages—specifically, how effectively this reference quality is maintained at the headset level (PC-HS and GT-HS).  The gap
from GT-PC to GT-HS is the fidelity cost introduced by bounded streaming
and on-device rendering.

Two key findings emerge from this analysis. First, local headset rendering closely replicates the host-rendered output in well-conditioned scenes. Specifically, PC-HS reaches 36.7 dB on office0 and 24.7–27.7 dB on the RealSense captures. This confirms that transmitting the model for local rendering—rather than streaming pre-rendered video frames—successfully preserves visual appearance across the pipeline. 

Second, the PC-HS metric decreases to 21.9–22.9 dB for large synthetic environments (office2, room0). This drop stems from field-of-view (FOV) mismatches rather than streaming artifacts. Because our evaluation metric spans the full, wide FOV of the headset, errors accumulate in the peripheral regions that were never captured in the training dataset. Consequently, the host and client architectures reconstruct these unobserved regions differently. When evaluating only the central region to match the dataset FOV, the PC-HS PSNR for office2 improves to 32 dB, aligning with its GT-PC baseline. 

Finally, the GT-HS column illustrates the end-to-end fidelity gap, with the lowest performance observed on lab1 (16.7 dB), which represents our most challenging capture. Regarding temporal stability, the system converges within 0.87–1.63 seconds following a distant teleportation, while post-convergence flicker remains negligible (under 0.20). For scenes characterized by larger SDF coverage and a smaller number of Gaussians (office0, office1), the flicker metric remains at zero. These metrics indicate that the transmitted model stabilizes rapidly and remains robust after viewpoint transitions.

\begin{figure*}[t]
  \centering
  \includegraphics[width=0.85\textwidth]{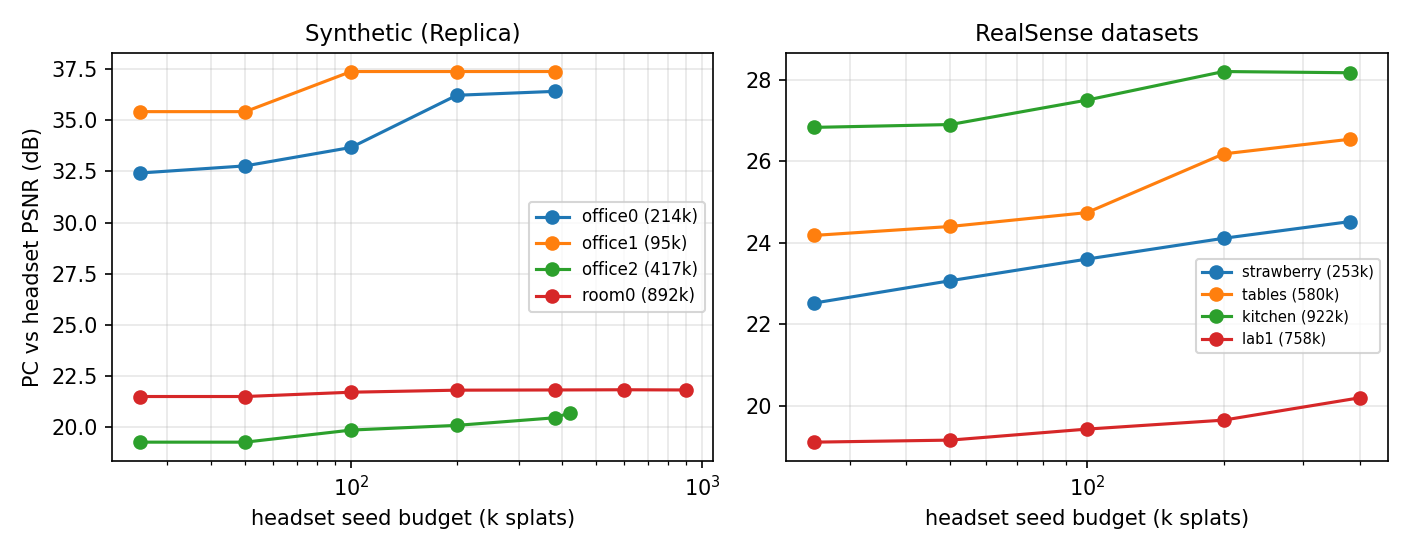}
  \caption{Client seed-budget ablation. PC-vs-headset PSNR as a function
  of the client resident-Gaussian budget (log axis), for the synthetic
  Replica scenes (left) and our RealSense datasets (right). Quality
  saturates near each scene's model size; on the large scenes the curve
  is nearly flat, showing that the per-viewpoint visible subset fits even
  a small budget.}
  \label{fig:budget}
\end{figure*}

\begin{figure*}[t]
  \centering
  \includegraphics[width=0.85\textwidth]{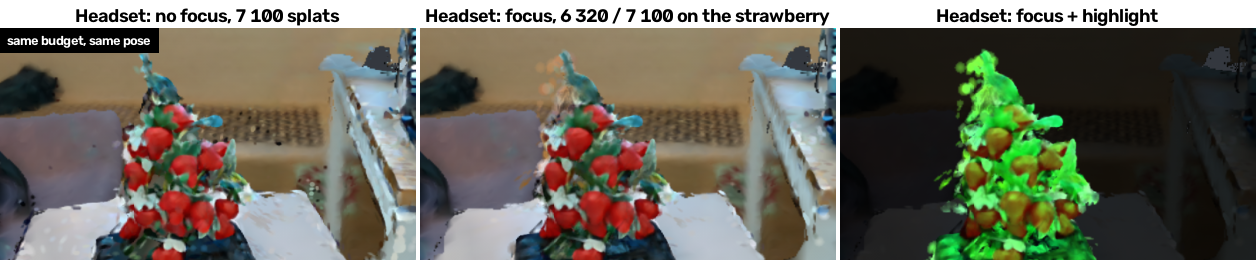}
  \caption{Semantic budget reallocation in the \texttt{strawberry} scene. Left: headset rendering without semantic focus using a 7,100-Gaussian budget. Center: semantic focus reallocates 6,320 of the 7,100 resident Gaussians (89\%) to the queried strawberry plant while retaining a coarse background representation. Right: the same focused representation with semantic highlighting enabled.}
  \label{fig:semantic}
  \vspace{-4mm}
\end{figure*}

\subsection{Client-Budget Ablation}

Fig.~\ref{fig:budget} reports PC-HS PSNR as a function of the
client resident-Gaussian budget for both scene groups on a log axis. Quality rises with budget until it saturates near the model size
(e.g. office1, 95k Gaussians, saturates by 100k), after which added
budget yields no gain. Critically, on the large scenes the curve is
nearly flat (office2 +1.4 dB, room0 +0.3 dB from 25k to the full
model): at any single viewpoint the visible subset fits even a small
budget, so the residency ceiling is set by the wide-FOV periphery rather
than by capacity. This supports the central design premise: a bounded
client model is sufficient for view-local fidelity, and the budget can be
spent selectively rather than uniformly.

\subsection{Semantic Budget Reallocation }

Tab.~\ref{tab:semantic} shows that an object query acts as a
deterministic allocation signal on the fixed client budget. Issuing a
class- or instance-level query concentrates 88–90\% of the resident
appearance Gaussians on the requested object while the total resident
count stays fixed, across objects of very different scale (a small
strawberry plant, a keyboard, a robot arm) and in both synthetic and
real scenes. The effect is a redistribution of the same budget, not an
increase: the object's share rises, and a thin background tier is
retained. As Fig.~\ref{fig:semantic} illustrates, the
queried object is rendered from streamed Gaussians while the remainder of
the room falls back to the coarser base representation.

The image-quality benefit of this reallocation is scene- and
viewpoint-dependent. At a near viewpoint that already fits the object,
the object-region PSNR difference between unfocused and focused
allocation is small (`strawberry`: 16.62 vs 16.93 dB over the object
mask), because view-dependent selection already retains most of the
object at that pose. The reallocation is therefore most valuable at
distant viewpoints or under tight budgets, where uniform selection would
otherwise drop object detail. We report semantic reallocation primarily
as a bounded model-state control mechanism and report image quality
separately from resident count.

\subsection{Standalone Rendering Performance}

We report frame \emph{time} rather than a single throughput number, because presented frame rate is bounded by the headset panel and the dominant source of frame-time variance is the publication of model updates, not local rendering. With the resident model fixed at 200k Gaussians and semantic reallocation and publication running concurrently, the native PICO 4 Ultra client renders each stereo frame at a steady-state GPU cost of approximately 9--10 ms (95th percentile) and a CPU cost of approximately 8 ms (95th percentile) on the office2/room0 scenes, rising to 11 ms on office1 where per-frame edit integration dominates. These steady-state costs sit below the frame budgets of both a 72 Hz panel
(13.9 ms) and a 90 Hz panel (11.1 ms).

\subsection{Qualitative results}\label{sec:qualitative}
\begin{figure*}[t]
    \centering
    \includegraphics[width=0.90\textwidth]{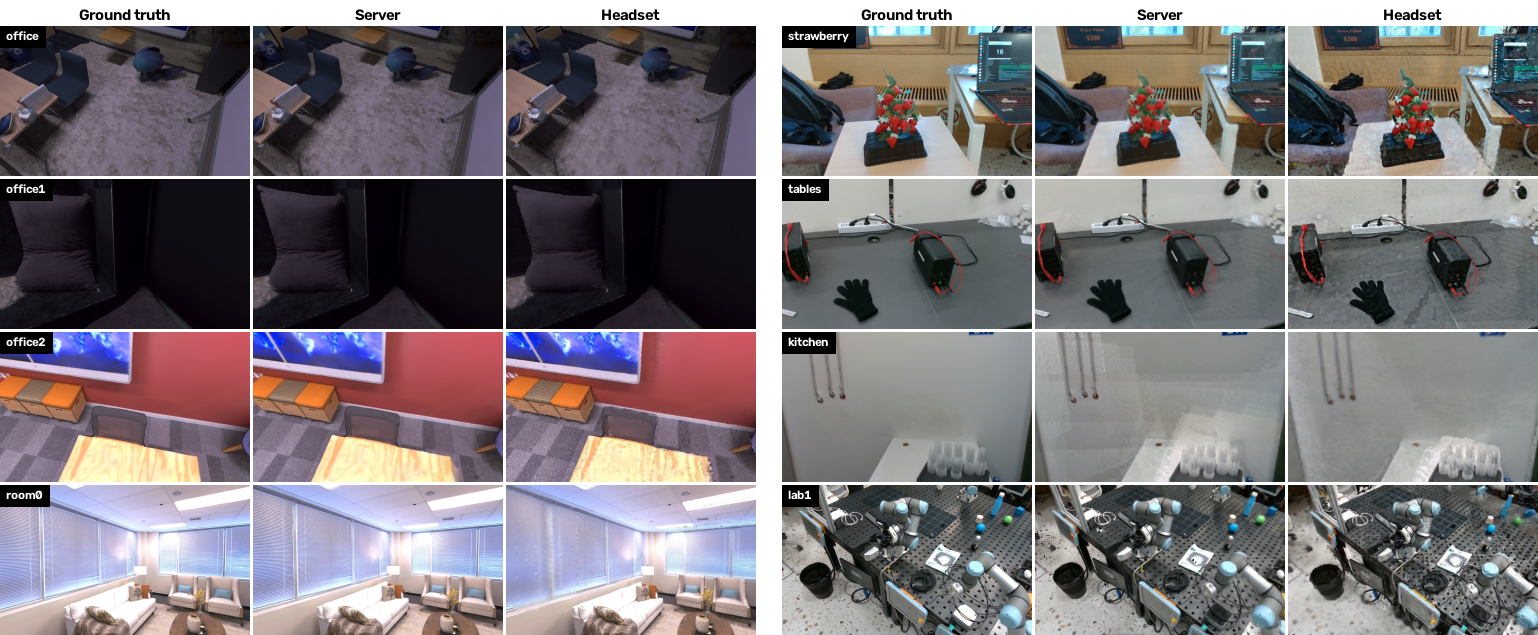}
    \caption{GT $\rightarrow$ server $\rightarrow$ headset across all eight
    evaluated scenes (GT, host render, and on-device headset render of the
    same camera). Left: synthetic Replica rooms; right: RealSense
    captures. The headset closely reproduces the host render on
    well-conditioned scenes; the larger residuals on the wide-FOV synthetic
    rooms concentrate in the periphery outside the captured frustum.}
    \label{fig:overview}
\end{figure*}

Fig.~\ref{fig:teaser} and Fig.~\ref{fig:overview} show the GT $\rightarrow$ server $\rightarrow$ headset chain
and semantic focus across scenes. Fig.~\ref{fig:stream} visualizes the continuous model update on the
headset during an \texttt{office2} fly-through: streamed Gaussians are
colour-coded as newly added (green), view-aware replaced (blue), or
removed (red) as the viewpoint moves. This shows that the bounded
resident model is incrementally added to, replaced, and pruned under head
motion while staying within its fixed capacity.

\section{Conclusion}
\label{sec:conclusion}

We presented GaussAnything, a standalone VR system for intent-conditioned allocation and staged publication of continuously evolving semantic Gaussian+SDF scenes. Across eight scenes, the native
PICO~4 Ultra client reproduced the host render on-device (up to 36.7\,dB
PC-HS) without remote frame streaming, and an explicit object query
redirected 88--90\% of the fixed client budget onto the requested object
without enlarging it. A budget ablation showed that a bounded resident
model is sufficient for view-local fidelity---quality saturates near the
model size and is nearly flat on large scenes---so the client budget can
be spent selectively rather than uniformly. The client renders each
stereo frame at roughly 10\,ms (GPU, 95th percentile), within the frame
budget of standard standalone panels, while model updates are published
and integrated incrementally under head motion.

These results highlight a broader distinction for immersive systems built on evolving reconstructions: deciding \emph{what} scene content should occupy limited client resources and deciding \emph{how} changes to that content should become visible are separate control problems. GaussAnything makes both explicit while preserving local head-coupled rendering, providing a practical foundation for standalone immersive applications in which scene representations continue to change in response to reconstruction and user intent.

\balance

\end{document}